\documentclass{article}
\usepackage{epsf}
\usepackage{graphicx}

\newcommand{\NP}{\rm {NP}}
\def\be{\begin{equation}}
\def\ee{\end{equation}}

\newcommand{\gev}{\rm{GeV}}
\newcommand{\tev}{\rm{TeV}}

\long\def\symbolfootnote[#1]#2{\begingroup%
\def\thefootnote{\fnsymbol{footnote}}\footnote[#1]{#2}\endgroup}

\newlength{\defbaselineskip}
\newcommand{\setlinespacing}[1]%
           {\setlength{\baselineskip}{#1 \defbaselineskip}}

\newcommand{\singlespacing}{\setlength{\baselineskip}{\defbaselineskip}}

\begin{document}


\begin{center} {\LARGE\textbf{Which fine-tuning arguments are fine?}}
\vskip 2em
{\large \bf Alexei Grinbaum} \\
{\it CEA-Saclay/LARSIM, 91191 Gif-sur-Yvette, France
\par Email alexei.grinbaum@cea.fr}
\vskip 1em \today
\end{center}
\vskip 1em
\begin{abstract}\noindent
The argument from naturalness is widely employed in contemporary
quantum field theory. Essentially a formalized aesthetic criterion,
it received a meaning in the debate on the Higgs mechanism, which
goes beyond aesthetics. We follow the history of technical
definitions of fine tuning at the scale of electroweak symmetry
breaking. It is argued that they give rise to a special
interpretation of probability, which we call Gedankenfrequency. By
extension of its original meaning, the argument from naturalness is
used to compare different models beyond the Standard Model. We show
that in this case naturalness cannot be defined objectively. Rather,
it functions as socio-historical heuristics in particle physics and
it contributes to the advent of a probabilistic version of Popper's
falsificationism.\end{abstract}


\section{Introduction}

Arguments from naturalness play an important role in particle
physics of the last 25 years. Gerard 't Hooft was the first to
introduce naturalness in this physical discipline, connecting it
with symmetry:
\begin{quote}
The naturalness criterion states that one such [dimensionless and
measured in units of the cut-off] parameter is allowed to be much
smaller than unity only if setting it to zero increases the symmetry
of the theory. If this does not happen, the theory is
unnatural.~\cite{tHooftNat}\end{quote}

Emphasized in this definition, the connection of naturalness with
symmetry could have provided a philosophical background for the
former based on the conceptual importance of the latter. Concerning
symmetry, since Plato and the 17th-century French debate between
Claude Perrault and Fran\c cois Blondel, two opposing views have
taken it to be, respectively, an expression of the aesthetic
imperative of beauty and a human-invented instrument for better
executing the job of the engineer. In turn, naturalness has both a
connection with beauty and a heuristic, road-mapping role in
science. Based upon 't Hooft's definition, it could have received a
double conceptual foundation similar to that of symmetry. But
history has chosen a more intriguing path. The original 't Hooft's
idea faded away behind the many facets of the actual use of
naturalness in particle physics.

Since the 1970s, the notion of naturalness has been gradually evolving
away from the connection with symmetry. In what physicists say about
its meaning one finds rare heuristic arguments as well as abundant
references to beauty: naturalness is an ``aesthetic
criterion''~\cite{AndersonCastano2}, a ``question of
aesthetics''~\cite{donoghue}, an ``aesthetic
choice''~\cite{AthronMiller}. Sometimes the aesthetical significance
of naturalness and the heuristic role are mixed together: ``the
sense of `aesthetic beauty' is a powerful guiding principle for
physicists''~\cite{GiudiceNat}.

One must not belittle the place of beauty in the scientist's
thinking. An intuitive aesthetic sense can be developed by the
practice of mathematical reasoning and it can then serve as a
thinking aid. In mathematics, once beauty and elegance have shown
the way to new discoveries, all the results must be rigorously
established through proof. In natural science, ``rational
beauty''~\cite{polking} can only be admired at the end of inquiry,
when we have established a sound scientific account in agreement
with nature. Einstein vividly supported this view early in his life,
saying that aesthetically motivated arguments ``may be valuable when
an \textit{already found} [his emphasis] truth needs to be
formulated in a final form, but fail almost always as heuristic
aids''~\cite{e2}. Used as a guide for discovering reality, aesthetic
arguments may indeed turn out to be extraordinarily fruitful as well
as completely misleading, and so for two reasons.

First, because the real universe is not just
beautiful: one can also discern in it futility~\cite{Weinberg79} or
inefficiency~\cite{DysonNYRB}. Nature is not what the American physicist
Karl Darrow thought she was, when he stated that it would be more ``elegant'' if there were
only two particles in the atomic nucleus~\cite{Darrow}. Perhaps
the most outspoken promoter of mathematical beauty in physics, Dirac
has many times been led by it into scientifically sterile
byways~\cite[chapter~14]{kragh}.
Thus beauty is not an exclusive characteristic of the results of science and should not be elevated
to a research imperative.

Second, because there is no necessary link
between beauty and empirically verified truth. The two notions are disconnected: the
beautiful may be false, and the true may be ugly. In spite of a long debate
on this topic between eminent physicists (e.g., see~\cite{chandra}), we maintain that beauty and truth,
as well as beauty and good, are distinct categories, in physics in particular. Aesthetic arguments are
a methodologically problematic and a potentially misleading
lighthouse on the road to sound science in the physical universe.

In Section~\ref{higgsection}, we review physics of the Higgs
mechanism and remind that the argument from naturalness often gives
the impression of being a perfectly normal scientific argument.
Section~\ref{measuresection} describes the many lines of development
of the concept of naturalness in particle physics. Among all
fine-tuning arguments, the valid one is neither anthropic
(Section~\ref{ontosection}) nor an argument from beauty. We argue in
Section~\ref{probapopper} that it involves a special interpretation
of probability and is meaningful only if naturalness in particle
physics is understood as heuristics. Practicality of being guided by
the considerations of fine tuning then stems not as much from
aesthetics as from the down-to-earth sociological factors, which
determine the physical theory's way of development. Naturalness can
be best warranted with the help of historical analysis.

\section{The Higgs mechanism}\label{higgsection}

The observed weak interaction is not locally gauge invariant and its
unification with electromagnetism must take it into account. To this
end, a mechanism must be introduced within any unified theory of
electroweak (EW) interactions to put the two interactions back on
unequal grounds. By offering one such mechanism the Standard Model
(SM) describes electroweak symmetry breaking quantitatively.
Invented in 1964 independently by several different groups, this
so-called Higgs mechanism builds on the fact that a massless
spin-one particle has two polarization states and a massive one has
three. Electroweak symmetry breaking produces a would-be Goldstone
boson, whose physical degree of freedom is absorbed by the massless
gauge boson. Number of polarization states of the latter then
increases from two to three and it becomes massive. Such massive
gauge bosons account for the absence of gauge symmetry in the
observed weak interaction.

This description was quickly recognized to be not very compelling
due to its lack of explanatory power~\cite{GiudiceNat,Rattazzi}.
Many physicists did not find important the conceptual problems of
the Higgs mechanism simply because they took it for no more than a
convenient, but temporary, solution of the problem of electroweak
symmetry breaking. Jean Iliopoulos said at the 1979 Einstein
Symposium: ``Several people believe, and I share this view, that the
Higgs scheme is a convenient parametrization of our ignorance
concerning the dynamics of spontaneous symmetry breaking, and
elementary scalar particles do not exist"~\cite{iliopoulos}. But,
with time, things have changed. Discovery of $W$ and $Z$ bosons and
a growing amount of electroweak precision data confirmed the ideas
of Weinberg and Salam. Not only is there today confidence in the
Standard Model, but it is clear that changing it ought to be
exceptionally difficult, due to an exceedingly large number of tests
with which any model beyond the Standard Model (BSM) must conform.
By 2004, Ed Wilson was completely assured: ``A claim that scalar
elementary particles were unlikely to occur in elementary particle
physics at currently measurable energies \dots makes no
sense''~\cite{wilson2}.

\begin{figure}[ht]
\begin{center}\epsfysize=4in \epsfbox{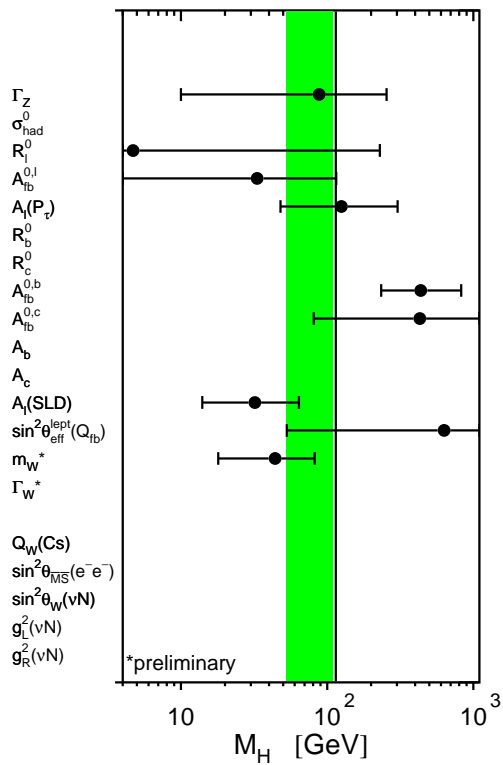} \caption{Values of
the Higgs mass extracted from different EW observables. The vertical
line is the direct LEP lower limit of 114 \gev. The average is shown
as a green band~\cite{lepewwg}.} \label{figHiggs}
\end{center}\end{figure}

The SM Higgs mechanism is a pleasingly economical solution for
breaking the electroweak symmetry. However, the global fit of the
electroweak precision data is consistent with the Standard Model
only in case one takes an average over all available experimental
data: then arises the usual prediction of a relatively light Higgs
$m_H<182\;\gev$~\cite{lepewwg}. Troubles occur, when one looks at
the details of the data: different ways of calculating the Higgs
mass $m_H$, based on distinct experimental measurements, lead to
incompatible predictions. Overlap between EW precision tests is less
than 2\% (Figure~\ref{figHiggs}).

For example, the value of the top quark mass extracted from EW data
(excluding direct Tevatron measurements) is
$m_t=178.9^{+11.7}_{-8.6}$~GeV, while the CDF/D0 result is $m_t =
172.6 \pm 0.8\mathrm{(stat)} \pm
1.1\mathrm{(syst)}$~\gev~\cite{top2008}. This discrepancy worsens
considerably the SM fit. Of more direct impact on the light Higgs
hypothesis is the observation that two most precise measurements of
the Weinberg angle $\sin^2\theta_W$ do not agree very well,
differing by more than $3\sigma$. The $b\bar b$ forward-backward
asymmetry $A_{fb}^{0,l}$ measured at LEP gives a large value of
$\sin^2\theta_W$, which leads to the prediction of a relatively
heavy Higgs with $m_H=420^{+420}_{-190}$~GeV. On the other hand, the
lepton left-right asymmetry $A_l$ measured at SLD (in agreement with
the leptonic asymmetries measured at LEP) gives a low value of
$\sin^2\theta_W$, corresponding to $m_H=31^{+33}_{-19}$~GeV, in
conflict with the lower limit $m_H>114$~GeV from direct LEP
searches~\cite{leph}. Moreover, the world average of the $W$ mass,
$m_W=80.392\pm 0.029$~GeV, is larger than the value extracted from a
SM fit, again requiring $m_H$ to be smaller than what is allowed by
the LEP Higgs searches~\cite{GiudiceF}.

For a physicist, inconsistency between the predictions of the Higgs
mass means that the argument in favor of the SM with a light Higgs
is ``less compelling''~\cite{GiudiceF}. What message exactly is
encoded in the 2\% overlap? Does this number correspond to some
probability? In what sense does the smallness of this value make the
SM Higgs less compelling?

\section{Measures of naturalness}\label{measuresection}

\subsection{Hierarchy problems}

The Standard Model suffers from a `big' hierarchy problem: in the
Lagrangian, the Higgs mass parameter $m_H^2$, which is related to
the physical mass by $m _h ^2 = -2 m _H ^2$, is affected by
incalculable cut-off dependent quantum corrections. Which\-ever new
theory, possibly including gravitation, replaces the Standard Model
above some energy scale $\Lambda_{\rm{NP}}$, one can expect the
Higgs mass parameter to be of the same size as (or bigger than) the
SM contribution computed with a cut-off scale $\Lambda_{\rm{NP}}$.
This way of estimating the size of the Higgs mass is made reasonable
by the analogy with the electromagnetic contribution to
$m_{\pi^+}^2-m_{\pi^0}^2$. The leading quantum correction  is then
expected to come from the top quark sector and is estimated to
be~\cite{Rattazzi} \be\delta m_H^2\sim
-\frac{3\lambda_t^2}{8\pi^2}\Lambda_{\rm{NP}}^2\,
.\label{quantcorr}\ee

This contribution is compatible with the allowed range of $m_h^2$
only if the cut-off is rather low \be \label{notuning}
\Lambda_{\NP}< 600 \times (\frac{m_h}{200\, \gev})\, \gev\, . \ee
Now, if the energy range of the SM validity is as low as $500$ \gev
~-- $1$ \tev, why did previous experiments not detect any deviation
from the SM predictions? Even though the center of mass energy of
these experiments was significantly lower than 1~\tev, still their
precision was high enough to make them sensitive to virtual effects
associated with a much higher scale.

To state it in other terms, note that effects from new physics at a
scale $\Lambda_{\NP}$ can in general be parametrized by adding to
the SM renormalizable Lagrangian a tower of  higher dimensional
local operators, with coefficients suppressed by suitable powers of
$\Lambda_{\NP}$: \be \label{effective} {\cal L}_{eff}^{\NP}
=\frac{1}{\Lambda_{NP}^2}\left \{c_1 (\bar e \gamma_\mu e)^2 +c_2
W_{\mu\nu}^I B^{\mu\nu}H^\dagger\tau_I H+\dots\right\}\,. \ee At the
leading order it is sufficient to consider only the operators of
lowest dimension, $d=6$. The lower bound on $\Lambda_{\NP}$ for each
individual operator ${\cal O}_i$, neglecting the effects of all the
others and normalizing $|c_i| = 1$, ranges between $2$ and $10$ TeV.
Turning several coefficients on at the same time does not
qualitatively change the result, unless parameters are
tuned~\cite{Rattazzi}. The interpretation of these results is that
if new physics beyond the SM affects electroweak observables at the
tree level, in which case $c_i\sim O(1)$, then the generic lower
bound on its threshold $\Lambda _{NP}$ is a few TeV. The tension
between this lower bound and eq.~(\ref{notuning}) defines what is
known as the `little' hierarchy problem.

The little hierarchy problem is apparently mild. But its behaviour
with respect to fine tuning is problematic. If fine tuning of order
$\epsilon$ is tolerated, then the bound in eq.~(\ref{notuning}) is
relaxed by a factor $1/\sqrt{\epsilon}$. The needed value of
$\epsilon$ grows quadratically with $\Lambda_{\NP}$, so that for
$\Lambda_{\NP} = 6$ TeV one needs to tune to 1 part in a hundred in
order to have $m_H=200$ GeV. The goal of this section is to make a
precise statement about the meaning of this fine-tuning problem.

\subsection{Standard definition}

The first modern meaning of naturalness is a reformulation of the
hierarchy problem. It arises from the fact that masses of scalar
particles are not protected against quantum corrections, and keeping
a hierarchical separation between the scale of EW symmetry breaking
and the Planck scale requires the existence of a mechanism that
would `naturally' explain this hierarchy. Although the difference in
scales is a dimensionless parameter much smaller than unity
($\frac{10^3\gev}{10^{19}\gev} = 10^{-16}$), setting it to zero in
accordance with 't Hooft's prescription is out of question because
gravity exists even if it is weak\symbolfootnote[1]{One exception
from this argument are models with large extra dimensions, where the
scale of gravity is different from $10^{19}$~\gev~\cite{ADD}.}. With
all its known problems, the Standard Model does not become more
symmetric in the hypothetical case where gravity is infinitely
weaker than the weak interaction. Therefore, 't Hooft's criterion
does not apply and naturalness needs a new definition.

According to Wilson ~\cite{susskind}, naturalness means that the
observable properties of a system should be stable against minute
variations of the fundamental parameters. This 1978 formulation
corresponds exactly to the lesson contained in the hierarchy
problem. It came at the end of a decade filled with debates on the
instability of the Higgs mass. In an article written at the end of
1970, Wilson had clearly stated his doubt that the Higgs mechanism
could be fundamental: ``It is interesting to note that there are no
weakly coupled scalar particles in nature; scalar particles are the
only kind of free particles whose mass term does not break either an
internal or a gauge symmetry. \dots Mass or symmetry-breaking terms
must be `protected' from large corrections at large momenta due to
various interactions (electromagnetic, weak, or strong). \dots This
requirement means that weak interactions cannot be mediated by
scalar particles''~\cite{wilson1}. After ten years of such and
similar doubts in the electroweak symmetry breaking through the
Higgs mechanism, the Standard Model was experimentally verified, and
little room remained for challenging its constitutive theoretical
components. If the hierarchy problem were to be tackled, the
Standard Model now had to be complemented rather than discarded.

In the years around 1980, supersymmetry advanced on the foreground
as a plausible extension of the problematic physics of electroweak
symmetry breaking in the Standard Model. Consequently, naturalness
began to be discussed in the context of supersymmetric models with
their enlarged content of particles and the new predicted phenomena,
e.g., in a seminal paper by Witten~\cite{witten}. As the number of
proposed supersymmetric extensions of the Standard Model increased,
a formalization of naturalness was needed to evaluate their
effectiveness in solving the big hierarchy problem. The first such
measure was defined in mid-1980s as a quantitative analogue of
Wilson's formulation.

Barbieri and Giudice looked at various realizations of low-energy
supersymmetric phenomenology arising from supergravitational
models~\cite{BarbieriGiudice,EllisEnquist}. They interpreted the
notion of naturalness by equating it with the sensitivity of the
electroweak symmetry breaking scale (instantiated as the $Z$-boson
mass $m_Z$) with respect to variations in model parameters. For a
general observable $O$ depending on parameters $p_i$ at point
$P^\prime$ this sensitivity is:
\begin{equation}\label{BG}
\Delta _{BG}(O;p_i) = \left| \frac{p_i}{O(p_i)} \frac{\partial
O(p_i)}{\partial p_i}\right|.
\end{equation}
Barbieri and Giudice then chose number 10 as a \textit{natural}
upper bound on $\Delta _{BG}$. The motivation was their subjective
belief that if the discrepancies between quantities were to be
natural, they must be less than of one order of magnitude. Yet, as
such, the choice of a number is arbitrary. In a different context
(discussing naturalness in semantic chains), Lewis shows that the
establishment of an endpoint of perfect naturalness is connected
with our own appreciation of what is ``not too
complicated''~\cite[p.~61]{LewisPlu}. The opinion in such matters
apparently can evolve: ten years after the Barbieri-Giudice
definition, when the experimental constraints on the leading BSM
candidate --- minimal supersymmetric standard model (MSSM) ---
became stronger, one had to require a fine tuning of 20 for the
model survival~\cite{CEP,barbieri}. Double the value of the old
endpoint, this new limit of naturalness also became accepted as
``reasonable''~\cite{kwok}.

Note that eq.~(\ref{BG}) only involves infinitesimal variations in
$p_i$. It follows that the Barbieri-Giudice definition gives a
measure of naturalness of a given model considered on its own,
independently of the rival models which differ in the values of
parameters but also pretend to solve the big hierarchy problem. This
definition has been used widely and has helped to sort out the
claims of different supersymmetric models about how well they
succeed in removing the big hierarchy problem of the Standard Model.
But it also failed to address a new set of issues in the
flourishing enterprise of model building.

\subsection{Naturalness in supersymmetric models}\label{NatSUSY}

In the late 1980s, BSM models began to be studied more thoroughly and a
multitude of their consequences became apparent, often unconnected
with the big hierarchy problem. Comparing this predicted
phenomenology with the growing ensemble of experimental data from
particle accelerators required a new notion of fine tuning. Now
naturalness must have encompassed many observables (and not just the
$Z$ mass). As a side effect of this evolution, the definitions of
naturalness no more considered only infinitesimal changes in
parameters, but a finite range of their values.

In spite of supersymmetry not being the only available solution of
the big hierarchy problem, a long line of studies have used fine
tuning to make guesses about the masses of sparticles. Early on, the
MSSM parameter space was scrutinized, later leaving the place to
that of NMSSM. In an article belonging to this current, de Carlos
and Casas~\cite{deCCasas}, who were critically reviewing an earlier
work which used the Barbieri-Giudice measure~\cite{RossRoberts},
realized that a measure of sensitivity need not always be a measure
of fine tuning. But they only concluded that one should take 20
rather than 10 as a numerical limit of natural $\Delta _{BG}$.

More radically, a newly defined measure appeared in 1994, when
Anderson and Casta\~{n}o refined the Barbieri-Giudice definition in
order to exclude such situations, i.e., when sensitivity is present
in a model for other reasons than fine
tuning~\cite{AndersonCastano}. They divide the Barbieri-Giudice
measure by its average value $\bar{\Delta}$ over some ``sensible''
range of parameters $p_i$:\begin{equation}\Delta _{AC} =
\frac{\Delta _{BG}}{\bar\Delta _{BG}} \label{AC}\end{equation} This
range can be specified by \textit{fiat} or can be chosen so as to
encompass all parameter values at which the model's experimentally
valid predictions remain unperturbed. Naturalness then can be
defined, in a slight modification of Wilson's language, as a
condition that observable properties of a system be ``not unusually
unstable'' against minute variations of the fundamental parameters.
The new word ``unusual'' implies comparison with the introduced
range of parameters and has a first-order conceptual importance.
Indeed, historically it has brought the meaning of the fine-tuning
argument in particle physics closer to probabilistic estimates based
on the anthropic reasoning.

That a range of parameters is involved in the definition of
naturalness means that parameter values in a particular model begin
to be seen as just one instantiation on the broader distribution of
\textit{possible} parameters. Anderson and Casta\~ no became the
first to connect naturalness to the ``likelihood'' of a given set of
Lagrangian parameters. They presupposed that there exist a way in
which ``we parametrize our assumptions about the likelihood
distribution of the theory's fundamental
parameters''~\cite{AndersonCastano3}. The range over which vary
parameters $p_i$ then arises as a mathematical representation of
such assumptions. Anderson and Casta\~ no were so led to consider a
class of identical models only differing in the values of
fundamental parameters, i.e., what we call today a landscape of
scenarii defined by the values of $p_i$. Distribution of parameters
over their allowed range was uniform and all values were considered
equally likely.

If Anderson and Casta\~ no were careful to speak about naturalness
only as likelihood of certain parameters, very soon did the word
`probability' enter the stage. Introduced by Strumia and his
co-authors, probability was not yet the probability of a particular
scenario seen on a landscape of many, but a mere inverse of the
Barbieri-Giudice measure of fine tuning. The latter was now
``supposed to measure, although in a rough way, the inverse
probability of an unnatural cancellation to occur''~\cite{barbieri}:
\begin{equation} P \sim \Delta _{BG}^{-1}.\end{equation} In a paper
discussing naturalness of the MSSM, Ciafalano and Strumia speak
about probability as a ``chance to obtain accidental cancellations''
in $M_Z$ ~\cite{CiaStrumia}. They attempt to demonstrate that the
choice of a particular limiting value of $\Delta _{BG}$ is no more
then a choice of a ``confidence limit on unprobable [sic]
calculations''. This is how probability as a degree of confidence,
i.e., in the Bayesian sense, made its way into particle physics.
Strumia goes on to suggest that probability could be normalized by
requiring that it be equal to 1 in the situations where ``we see
nothing unnatural''. What this phrase means precisely is left to our
subjectivity. However, the normalization problem is very important:
its difficulty lies with the fact that most attempts to rigorously
define parameter space lead to non-normalizable solutions, so that
it is impossible to define the ratios between regions of these
spaces~\cite{mcgrew}. Thus Strumia's use of `probability' is
metaphoric. This probably was the reason why, previously, Anderson
and Casta\~ no had avoided this term and had only spoken about
`likelihood'.

The originally metaphoric phrase ``roughly speaking, $\Delta ^{-1}
_{BG}$ measures the probability of a cancellation'' proved popular
(see, e.g.,~\cite{ceh1}). It was used by Giusti \textit{et al.},
when they variously spoke about ``naturalness probability'' or
``naturalness distribution probability''~\cite{Giusti}. This line of
thought refers to probability because it needs a justification for
doing a Monte Carlo calculation of ``how frequently numerical
accidents can make the $Z$ boson sufficiently lighter than the
unobserved supersymmetric particles''. It is remarkable that,
although Bayesian in its roots (Monte Carlo being a Bayesian
method), probability is seen here as the frequency of an event
occurring only in the thought experiments, performed by an agent who
imagines worlds with different values of the parameters of
supersymmetry. This \textit{Gedankenfrequenz} interpretation of
probability becomes typical for a group of papers on fine tuning in
supersymmetric models. Although frequentist in its formulation, it
is a variant of the Bayesian point of view because it relies on the
subjective assignment of priors, which corresponds to Anderson's and
Casta\~ no's ``way in which we parametrize our assumptions''.
Initially the agent's freedom to give a value to the prior
probability is limited by the boundaries of the allowed region of
parameter space. Once in the allowed region, stategies vary. On the
one hand, Giusti \textit{et al.} propose to choose values randomly
and use them in a calculation which leads to assigning a Bayesian
level of confidence to the sets of parameters. On the other hand,
among many articles using the Markov Chain Monte Carlo (MCMC)
procedure for MSSM, one finds other choices of priors, such as
``naturalness-favouring prior''~\cite{Allanach} or ``theoretical
probability of a state of nature''~\cite{Cabrera}.

Resulting in what they called `LHC forecasts', these Bayesian
studies make ``though reasonable, rather arbitrary'' predictions
about future experiments. It is important that this approach has
paved the way to understanding Strumia's metaphoric probability in
the statistical sense. When ten years later Casas \textit{et al.}
will be comparing definition~(\ref{deltasq}) with definition
(\ref{deltamax}), they will speak about ``the statistical meaning''
of fine tuning~\cite{ceh4}.

\subsection{Naturalness in model comparison}

Defining naturalness with the help of a finite range of parameters
corresponding to different model-building scenarii became a dominant
trend. Particle physics was now seen as consisting of
scenarii~\cite{AthronMiller,ceh3,ceh4}. Naturalness was redefined in
this new language: it became a measure of ``how atypical'' are
certain physical scenarii~\cite{AthronMiller}. If the use of fine
tuning had previously been limited to emphasizing the problems of a
particular model, many physicists state after the year 2000 that
naturalness is used to compare models.

Anderson and Casta{\~ n}o modified the Barbieri-Giudice measure,
eq.~(\ref{AC}) instead of eq.~(\ref{BG}), because of the problem of
global sensitivity. Athron and Miller went further to consider
models with several tuned observables as well as finite variations
of parameters~\cite{AthronMiller}. Parameters themselves are no more
required to be uniformly distributed over the considered range of
parameter space. To give a quantifiable version of this larger
notion, Athron and Miller speak about ``generic'' scenarii and
``typical'' volumes of parameter space formed by ``similar''
scenarii. Introduced in the first modifications of the
Barbieri-Giudice measure as a finite range of parameters,
exploration of the larger parameter space far from the point
$P^\prime$ reaches here its apogee.

To define `similar' and `typical', Athron and Miller claim in
opposition to Anderson and Casta\~ no that the definitions must be
``chosen to fit to the type of problem one is considering''. They
argue that a typical volume of parameter space cannot be the
Anderson-Casta\~ no average of volumes $G$ throughout the whole
parameter space, $\langle G\rangle$, for it would depend only on how
far the parameters are from some ``hypothesized upper limits on
their values''. For example, an observable $O$ which depends on a
parameter $p$ according to $O=\alpha p$, will display fine tuning
for small values of $p$ if one chooses the maximum possible value of
$p$ to be large. In the Anderson-Casta\~ no approach, upper limits
on parameters arise from the requirement that the model's meaningful
predictions be preserved. For Athron and Miller this is too generic.

To fit the choice to particular cases, they introduce similar
scenarii defined by a ``sensible'' choice of how far numerically the
observable value may deviate from a given one. Let $F$ be the volume
of dimensionless variations in the parameters over some arbitrary
range $\left[ a,b\right]$ around point $P^\prime$ and $G$ be the
volume in which dimensionless variations of the observable fall into
the same range:
\begin{equation}
a \leq \frac{p_i \left( P\right)}{p_i \left( P^\prime \right)} \leq
b,\quad a \leq \frac{O_j \left( \{ p_i \left( P \right)\}\right)
}{O_j \left( \{ p_i \left( P^\prime \right) \} \right)} \leq b.
\end{equation}
In their MSSM calculation Athron and Miller use $a=b=0.1$ claiming
that this $10\%$ threshold amounts to not encountering a
``dramatically different'' physics. The measure of fine tuning then
is
\begin{equation}\label{AM}\Delta _{AM} = \frac{F}{G}.\end{equation}

This measure can be applied straightforwardly in the case of a
single observable like the $Z$ mass, but it can also be applied to
compare the tuning between different observables. In the latter case
$F$ and $G$ are volumes in the multi-dimensional spaces of,
respectively, parameters and observables. The former space is not
new, for the notion of naturalness defined by many parameters dates
back to Barbieri and Giudice:
\begin{equation}\label{deltamax}
\Delta = \max _i \{\Delta_{BG} {\left( p_i \right)}\}.
\end{equation}
Alternative variants have also been proposed, such
as~\cite{ceh2,ceh3}
\begin{equation}\label{deltasq}
\Delta = \sqrt{\sum _i \Delta_{BG} \left( p_i \right) }.
\end{equation}
On the contrary, introducing a multi-dimensional space of
observables is a novelty.

With the development of model building it became clear that the big
hierarchy problem was not the only fine tuning to be found. Many
experimental parameters were measured constraining the values of
parameters in BSM models, such as quark masses, the strong coupling
constant, the anomalous magnetic moment of the muon, the relic
density of thermal dark matter, smallness of flavor violation,
non-observation of sparticles below certain thresholds and so forth.
Fine tunings produced by these measurements are ``morally
similar''~\cite{SchusterToro} to the fine tuning from $m_Z$. A
variety of heuristics are then possible. One can consider the most
constraining fine tuning or some form of the average of many
tunings. Motivations from different tunings may not be equally
``compelling'' and less remarkable parameters may be therefore
discarded.

Still the Anderson-Casta\~ no problem of upper limits on $p$ cannot
be avoided even if one defines similar scenarii independently.
Athron and Miller wish to maintain decorrelated tunings and to vary
each observable without regard for others. Individual contributions
to volume $G$ are then made with no concern for contributions from
other observables. At this point Athron and Miller realize that
observables can only be compared if $\Delta _{AM}$ is normalized. To
do so, they are forced to reintroduce the Anderson-Casta\~ no
average value~(\ref{AC}):
\begin{equation}\label{AMaverage}
\hat{\Delta} _{AM} = \frac{1}{\bar\Delta} \frac{F}{G},
\end{equation}
which relies on the knowledge of the total allowed range of
parameters in a particular model. The hypothesized upper limit of
this range determines how compelling the naturalness argument for
new physics will be. The same normalization procedure is essential
if one wants to use fine tuning to compare different models.

Although it appears in the literature as an incremental refinement
of the original Wilson's idea through the work of Barbieri, Giudice,
Anderson, Casta\~ no and others, the Athron-Miller notion of
naturalness lies very far from Wilson's. Naturalness has become a
statistical measure of how atypical is a particular scenario. It is
now tempting to use the numerical value of fine tuning to set off
several scenarii against each other: the least tuned scenario is to
be preferred, where `preferred' is understood in a practical,
heuristic sense of model building. In practice, not only similar (in
the Athron-Miller sense) scenarii are compared according to the
value of their fine tuning, but models predicting completely
different physics are sometimes brought into a competition against
each other on the basis of their naturalness. On the one hand, one
reads that:
\begin{quote}
    The focus point region of mSUGRA model is \textit{especially compelling}
    in that heavy scalar masses can co-exist with low fine-tuning\ldots~\cite[our emphasis]{baer}

    We \ldots find \textit{preferable} ratios which reduce the degree of fine tuning.~\cite[our emphasis]{abe}
\end{quote}
On the other hand, such claims are mixed with assertions going
beyond the applicability of the Anderson-Casta\~ no or even the
Athron-Miller definitions:
\begin{quote}
    Some existing models\ldots are not \textit{elevated} to the \textit{position} of supersymmetric
    standard models by the community. That may be because they involve fine-tunings\ldots~\cite[our emphasis]{bine}

    In order to be \textit{competitive} with supersymmetry, Little Higgs models
    should not worsen the MSSM performance [in terms of the degree of fine tuning].
    Fine tuning much higher than the one associated to the Little
    Hierarchy problem of the SM \ldots or than that of supersymmetric models \ldots is a serious
    drawback.~\cite[our emphasis]{ceh3}

    \ldots the fine-tuning \textit{price} of LEP\ldots~\cite[our emphasis]{CEP,barbieri}
\end{quote}
Comparing altogether different models by confronting the numbers,
e.g., being tuned at 1\% against being tuned at 10\%, is meaningless
unless one of two conditions is met: either the two models can be
put in the common parameter space and the Athron-Miller definition
is to be used, or conclusions drawn from such comparison are
employed in a particular way. Because they cannot be said to bear on
truth value of the models, they can be understood as shaping the
historic and sociological competition between otherwise
uncommensurable models.

\section{Ontological interpretation}\label{ontosection}

\subsection{The anthropic connection}

The naturalness problem was one of the factors that gave rise to
theories beyond the Standard Model. Since the 1970s, the SM began to be
viewed as an approximation to some future fundamental theory, i.e.,
an effective field theory (EFT) valid up to some limit $\Lambda
_{NP}$. The fundamental theory may involve gravity, and the SM would
then become its low-energy limit. The EFT approach relies crucially
on the assumption of decoupling between energy scales and the
possibility to encode such a decoupling in a few modified constants
of the field-theoretic perturbation series. This connects EFT with
naturalness.

Understood as a hierarchy problem, naturalness is the measure of
stability against higher-order corrections in the perturbation
series. If the higher-order corrections were important, it would
invalidate the use of the perturbation expansion and, together with
it, the EFT method. ``If the experiments at the LHC find no new
phenomena linked to the \tev ~scale, the naturalness criterion would
fail and the explanation of the hierarchy between electroweak and
gravitational scales would be beyond the reach of effective field
theories. But if new particles at the TeV scale are indeed
discovered, it will be a triumph for our understanding of physics in
terms of symmetries and effective field
theories''~\cite{GiudiceNat}.

If low-energy models (e.g., MSSM) are EFTs with respect to some
unified theory involving gravitation (e.g., supersymmetric models of
gravity), it is possible to speak within one and the same theory
about the fine tuning of low-energy observables (like $m_Z$) as well
as about the fine tuning of the cosmological constant. Thus fine
tuning in particle physics and fine tuning is cosmology become
connected. While the latter tuning has a long tradition of been
interpreted anthropically, it is through this connection that the
former tuning acquires a tint of anthropic meaning.

Introduction of the range of parameter values in the definitions of
naturalness, eqs. (\ref{AC}), (\ref{AM}), and (\ref{AMaverage}),
pushes one in the direction of the many-worlds ontology. If every
value from the range of parameters is realized in some world, one
can justify the fine tuning argument as a probability distribution
corresponding to our chances to find ourselves in one of these
ontologically real worlds. This interpretation seems totally
fictitious, but it is the one shared intuitively by many physicists,
particularly string theorists and cosmologists~\cite{unimulti}. It
inserts the fine-tuning argument in a larger class of anthropic
arguments based on the many-worlds reasoning.

The argument goes as follows. 1$^\circ$, establish that the
descriptions of worlds with different values of parameters are
mathematically consistent and not precluded by the theory.
2$^\circ$, establish that such worlds really exist. For this, refer
to Gell-Mann's ``totalitarian principle'', requiring that anything
which is not prohibited be compulsory~\cite{bilaniuk}.
Alternatively, refer to what Dirac called ``Eddington's principle of
identification'', that is, asserting the realist interpretation of
mathematical quantities as physical entities~\cite{dirac31C}. Or
extrapolate to all physics Peierls's position that ``in quantum
electrodynamics one has always succeeded with the principle that the
effects, for which one does not obtain diverging results, also
correspond to reality''~\cite{peierls}. 3$^\circ$, establish that
among all possible worlds those containing highly fine-tuned models
are statistically rare, for their probability is defined by the
inverse fine tuning. Indeed, the definition of ``unnatural'' was so
chosen that, compared to the full number of worlds, the proportion
of unnatural worlds is necessarily tiny. 4$^\circ$, conclude that if
we evaluate our chances to be in such a world, the resulting
probability must be low.

This argument can, and has been, criticized at every step from 1 to
4. For example, depending on the concrete variety of the anthropic
argument, the pronoun `we' (step 4) refers either to intelligent
beings, or worlds with carbon-based life, or else worlds with
complex chemical elements and so forth. Logically, everything
happens as if there were a choice-making meta-agent with access to
reason. But with an ontological interpretation of worlds, this
meta-agent becomes a super-agent with a power of action extending
over the many worlds, who blindly decides to put us in a world of
her choice. The existence of such super-agent is of course
metaphysical. Yet it warrants the posture that, provided our task to
predict in which world we shall end up, we cannot fare better than
guess it probabilistically, by taking the inverse of the fine tuning
measure $\Delta$.

The specific problem of the anthropic argument in particle physics, emphasized in Section~\ref{measuresection}, is
that the `full number of worlds' (step 3) can be only
defined arbitrarily. Upper limits of the range of parameter values
have to be set by \textit{fiat}. If one goes in this too far, it would preclude
some worlds from existing without a contradiction with theory or data, thus violating the requirement
at step 1. But how far one can go in parameter variations and keep the premises of step 1 intact is not obvious.
At the first sight, this difficulty may seem unremarkable and the anthropic argument would seem a valid
implication. Inconspicuity of the limit problem, which is often left unmentioned, is similar to the general disregard among
physicists of the frequent use of `probability' in a metaphoric, formally indefensible sense.
Yet if the story of naturalness in particle physics teaches a clear lesson about anthropic reasoning, it is about how to
show its arbitrariness on a concrete example.

\subsection{Counterfactuals}

The fine-tuning argument shares with a larger class of anthropic
arguments a twofold logical nature: these arguments can either be
formulated in purely indicative terms or by using counterfactuals.
The first kind of formulations, using only indicative terms, are
typically employed by opponents of the anthropic
principle~\cite{SmolinAnthPr}. They mean to dissolve the apparent
explanatory power of the argument by rewording it in terms of facts
and of the laws of inference in classic Boolean logic. Devoid of the
counterfactual, the anthropic argument indeed becomes trivial.

The second kind of logic involving explicit counterfactuals is more
common. Anthropic arguments take the form of statements such as `If
parameters were different then intelligent life would not have
existed''; or `If parameters were different then complex chemistry
would not have existed'; or `If parameters were different then
carbon-based life would not be possible'. The principal question is to find out
whether such statements are explanatory, and if yes, then in what sense.
Answers to it typically involve a detailed analysis of the counterfactual semantics.
We only note here that there exists a physical, but no less fundamental than any logical, problem of
validity and applicability of a counterfactually formulated argument.

Counterfactuals in physics have been discussed at least since the
Einstein, Podolsky and Rosen paper about quantum mechanics in
1935~\cite{EPR}. The key point in the EPR argument is in the
wording: ``If\ldots we had chosen another quantity\ldots we should
have obtained\ldots''. The Kochen-Specker theorem and Specker's
discussion of counterfactuals in 1960 placing them in the context of
medieval scholastic philosophy were the starting point of a heated
debate on the use of counterfactuals in quantum mechanics (for
recent reviews see~\cite{Vaidman,Svozil}). Peres formulated perhaps
clearest statements about the post-Bell-theorem status of
counterfactuals:
\begin{quote}
The discussion involves a comparison of the results of experiments
which were actually performed, with those of hypothetical
experiments which could have been performed but were not. It is
shown that it is \textit{impossible to imagine} the latter results
in a way compatible with (a) the results of the actually performed
experiments, (b) long-range separability of results of individual
measurements, and (c) quantum mechanics. \ldots

There are two possible attitudes in the face of these results. One
is to say that it is illegitimate to speculate about unperformed
experiments. In brief ``Thou shalt not think.'' Physics is then free
from many epistemological difficulties.\ldots Alternatively, for
those who cannot refrain from thinking, we can abandon the
assumption that the results of measurements by $A$ are independent
of what is being done by $B$. \ldots Bell's theorem tells us that
such a separation is impossible for individual experiments, although
it still holds for averages.~\cite{Peres78}
\end{quote}

The debate in quantum mechanics shows that the applicability of
Boolean logic to statements about physical observables should not
taken for granted in any branch of physics, especially those based
on quantum mechanics. Quantum field theory is one. Simply, its focus
has stayed with technical feats for so long that conceptual issues
about measurement, inherited from quantum mechanics, have been
neglected. The tendency has prevailed to assign values to unobserved
parameters in experimental settings which cannot be realized in
principle, e.g. in the case of \textit{Gedankenfrequenz}.

Admittedly, even if the counterfactual in the fine-tuning argument
in particle physics bears on physical parameters in the worlds
impossible to observe, this does not lead to a direct contradiction
with quantum mechanical theorems, for quantum mechanics deals with
normalized probability spaces and Hermitian observables. It
nonetheless remains true that the logic of anthropic arguments runs
counter to the trend warranted by the lessons from quantum
mechanics. Speculation about unperformed experiments is illegitimate
not only in the case of unrealized measurements of Hermitian
operators, but in a more general sense: it is unsound to extend to
unperformed experiments in unrealized worlds the Boolean logical
structure allowing us to say that physical constants in those worlds
have definite values.

This line of critique resonates with Bohr's answer to Professor
H{\o}ffding when the latter asked him and Heisenberg during a
discussion at the University of Copenhagen: ``Where can the electron
be said to be in its travel from the point of entry to the point of
detection?'' Bohr replied: ''To be? What does it mean \textit{to
be}?''~\cite[p.~18-19]{Wheeler} The fine-tuning argument in particle
physics, as well as anthropic arguments involving the cosmological
constant, employ counterfactuals that contain the verb `to be' in
the conditional. What it means that a world which is referred to in
this conditional, \textit{had been}, \textit{was} or \textit{is},
would have been unclear for Bohr. He was greatly concerned with the
meaning of utterances, famously claiming that ``physics is what we
can say about physics''~\cite[p.~16]{Wheeler}. In the case of fine
tuning this claim may be understood as a radical warning to all
those who interpret fine tuning ontologically.

\section{Probabilistic falsification}\label{probapopper}

\subsection{Interpretation of probability}

Casas \textit{et al.} consider two tunings of two different
observables and propose that ``since $\Delta$ and $\Delta
^{(\lambda)}$ represent independent inverse probabilities, they
should be multiplied to estimate the total fine tuning $\Delta \cdot
\Delta ^{(\lambda)}$ in the model''~\cite{ceh3}. This is clear
evidence of the statistical meaning of naturalness, shared by many
particle physicists since the work of Ciafaloni and
Strumia~\cite{CiaStrumia}. We argued in Section~\ref{NatSUSY} that
the notion of probability must be interpreted here in a peculiar way
combining a frequentist approach with Bayesianism. Frequency is
\textit{Gedankenfrequenz}, because one counts the number of
particular occurrences in the class of imaginary untestable
numerical scenarios, instantiated as points in the parameter space.
The Bayesian component arises in the form of degree of confidence,
for one is concerned with our current ignorance of the true value of
a parameter, which we believe to be measured in the future. When a
parameter has already been measured, even outspoken proponents of
the statistical meaning of naturalness admit that ``assigning to it
a probability can be misleading''~\cite{ceh3}. Therefore, the
fine-tuning argument is to be understood as a bet on our future
state of knowledge, and it loses all meaning at the moment of actual
measurement. \textit{Hic et nunc} the future state of knowledge does
not exist, and the bet is subjective. However, in our mind it
does exist, and in this mental reality naturalness can be
interpreted as frequency.

Two conditions must be met for one to have the ability, by way of
naturalness, of making bets on future physics. First, experimental
data must lack dramatically, so that uncertainty be complete as for
which model better describes reality. This is indeed the case in
particle physics. Second, we must hold a belief that in the future,
hopefully soon enough, this veil of uncertainty will be lifted.
Moreover, we ought to take it for granted that the lifting of the
veil will unambiguously determine which of our current models is
right and which is wrong. Subjective bets make sense only if the
future state is a choice between the available alternatives.
Physicists using the fine tuning argument hold such a belief indeed.
If they refer to naturalness, they assume full confidence that the
true description of reality will be picked out of the current models,
once and for all, at the LHC or later
experiments.

As in the general case of probabilistic reasoning in a situation of
uncertainty (see, e.g.,~\cite{Nickerson}), the fine tuning argument
is the last resort when no scientific explanation can be provided.
Psychologically, it is very difficult to resist the temptation to
make a ``statistical guess''~\cite{ceh3} at the future state of
knowledge. Although naturalness provides guidance but adds nothing to scientific truth,
accepting its irrelevance and ``living with the existence of fine
tuning''~\cite{donoghue} is a hard way of life. This temptation
though is not completely unfamiliar as we already live in a world
with many fine tunings, for example:
\begin{quote}\begin{itemize}
\item The apparent angular size of the Moon is the
same as the angular size of the Sun within 2.5\%.
\item The recount of the US presidential election results in Florida in 2000 had the official result
of 2913321 Republican vs. 2913144 Democratic votes, with the ratio
equal to 1.000061, i.e., fine-tuned to one with the precision of
0.006\%.
\item The ratio of 987654321 to 123456789 is equal to 8.000000073, i.e., eight with the precision of $10^{-8}$.
In this case, unlike in the previous two which are
coincidences, there is a `hidden' principle in number theory, responsible for the large amount of fine tuning.~\cite{lands}
\end{itemize}\end{quote}

By the analogy with the last example, if we believe that a `hidden'
new principle in particle physics will be uncovered (and we believe
it if we place bets on the future state of knowledge), running a
competition between models by comparing their amount of fine tuning
may seem to bring us closer to uncovering the principle. However,
to adhere to this idea would be logically and methodologically incorrect, for the principles
of Nature, both known and unknown, are unique --- i.e., unstatistical --- and independent of our
will. Then, naturalness can only serve to satisfy a human
psychological urge. We use it to please the senses by setting off
the models in a beauty contest. Such a disappointing verdict for
naturalness would indeed sound grim and gloomy, had it not been for
its different, and a less nebulous, function.

\subsection{Naturalness as heuristics}

Karl Popper's falsification, which took much inspiration from the pretense to adequately describe the
methodology of high-energy physics, relies on the assumption that physical
experiment can rule out definitively certain predictions made within
theoretical models. If this is the case, then the models, or at
least such elements of these models that are directly responsible for the unfulfilled
predictions, do not describe physical reality, and are false.

Popperian methodology depends critically on the possibility to
interpret experimental data. If the findings are not conclusive,
models cannot be falsified in the original sense of Popper's. Yet in
particle physics of the last 25 years experimental findings have not been conclusive.
While the power of particle accelerators grows and their exploratory
capacity continues to be gradually augmented, no recent accelerator experiment has falsified a theoretical model, even if,
in accordance with the falsification doctrine, the
predicted phenomena had not been observed. This is chiefly because experiments at particle accelerators, as
well as the gathered cosmological data, are so complex that one is
unable to set up a unique correspondence between data and predictions made within theoretical models.
We often ignore if we already possess a name or a theory for what we have observed.
Unambiguous falsification of the models in particle physics is therefore
impossible. At best, experimental findings suggest that certain
predictions, while not completely ruled out, are rather difficult to
sustain as open possibilities.

Departing from the original Popper's view, methodology of particle physics mutates thus
into its probabilistic
version. Complex experiments at the accelerators leave any model with a chance to die and a chance to
survive, but never act as definitive model murderers.
Notwithstanding, a model can still die: not because it was falsified, but merely for falling out of
fashion.

The rise and fall of theories and models in contemporary
particle physics is more a matter of a partly circumstantial history
than subject to a rigorous epistemology. Influence
of the sociological factors can be decisive, e.g., the choice at the leading
universities of professors with a particular taste in physics, or
the abrupt reversals between fashionable and worn-out lines of
research. The argument from naturalness is a powerful instrument for
influencing the development of particle physics, for, in virtue
of having the form of a normal scientific argument, it can speak to the
scientist. Indeed, arbitrariness of the measure of naturalness used for comparing different models
is disguised. On the surface, one only encounters a presumably legitimate
comparison of numbers, which does not bear any sign of the underlying problematic
choice of the limiting range of parameters in the parameter space.

Those who are the first to fix the arbitrary convention of what is natural and
what is not, exercise significant influence over those who will
follow later. Wilson, Barbieri and Giudice did so. In particular,
Barbieri and Giudice gave a mathematical definition of fine tuning,
providing a definite form to what had only been a vague feeling of
aesthetical unease. Ever since the 1990s, their work has been
turned, albeit usually in the hands of others, into a powerful sociological instrument.

Imagine two models which theoretically explain away the big
hierarchy problem, meanwhile no experimental measurements can be
made to distinguish between them. The only possible competition
between the models is based on purely mathematical criteria such as
the numerical value of fine tuning. To know which model will win in
the course of history, provided that experimentalists are unable to
settle this question, one can only make guesses. And to make a
plausible bet in this uncertain future-oriented competition, it may
be helpful to use the heuristics of naturalness.

Now imagine that in the future the current argument from naturalness
is sociologically overrun by something else. No matter what the
modification will be, it is only likely that the argument accepted
by the scientific community will mutate into another argument
accepted by the scientific community. That fine tuning would be
altogether proclaimed irrelevant or invalid, can hardly be
envisaged. Influence of the naturalness heuristic initiated in the
1980s cannot be erased. Indeed, one such modification happened
around 2000, when, without having previously been a scientific
argument for model comparison, the argument from naturalness grew
into presenting itself as such, and at the same time continuity was
forcefully proclaimed with the original notion.

Trends in the model building in particle physics were formed by the
natural\-ness-based heuristics, which is instrumental in dismissing
``unnatural'' theories and could yet lead to a ``more complete
model'' explaining stability of the weak
scale~\cite{AndersonCastano}. Physicists sometimes see clearly this
mutation of the fine-tuning heuristics. For example, Bin\' etruy
\textit{et al.} warn that
\begin{quote}
The [fine-tuning] approach should be treated as providing guidance and should not
be used for conclusions such as ``the heterotic string theory on an
orbifold is $3.2\sigma$ better at fitting data than a Type I
theory...''~\cite{bine}
\end{quote}
But even if such warnings are heard, and the direct judgment of the
kind ``one model is better than another'' avoided, the sociological
heuristics is still at work. One its manifestation is that working
physicists will turn away from highly tuned models because their
faith in them has been lost. This is connected with the Bayesian meaning of
naturalness as ``degree of confidence''.

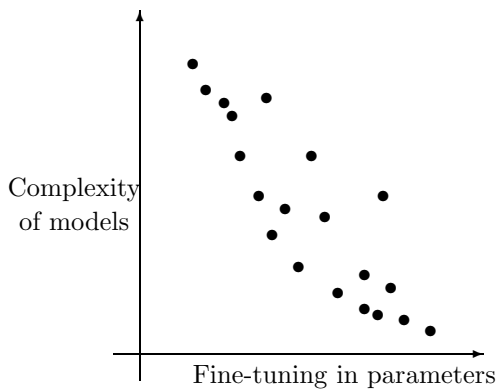
\begin{figure}[htbp] 
   \centering
\begin{picture}(200,150)(0,0)
\put(40,15){\vector(1,0){140}} \put(50,5){\vector(0,1){140}}
\put(70,4){Fine-tuning in parameters} \put(0,75){Complexity}
\put(1,62){ of models} \put(160,24){\circle*{4}}
\put(150,28){\circle*{4}} \put(145,40){\circle*{4}}
\put(142,75){\circle*{4}} \put(140,30){\circle*{4}}
\put(135,45){\circle*{4}} \put(135,32){\circle*{4}}
\put(125,38){\circle*{4}} \put(120,67){\circle*{4}}
\put(115,90){\circle*{4}} \put(110,48){\circle*{4}}
\put(105,70){\circle*{4}} \put(100,60){\circle*{4}}
\put(95,75){\circle*{4}} \put(88,90){\circle*{4}}
\put(98,112){\circle*{4}} \put(85,105){\circle*{4}}
\put(82,110){\circle*{4}} \put(75,115){\circle*{4}}
\put(70,125){\circle*{4}}
\end{picture}
   \caption{Schematic graph of fine tuning versus model complexity in the space of models beyond SM~\cite{cheng}.}
   \label{ftuning}
\end{figure}

Another manifestation is that the heuristics of naturalness has so
influenced model building that no simple model without significant
fine tuning remains in the valid model space (Figure~\ref{ftuning}).
Unnaturalness of simpler models led to the development of more
complicated ones, which are allegedly less tuned. Even if at the end of
the day such more complex models often turn out to be as tuned as simpler
ones (e.g., see~\cite{SchusterToro}), the sociological and historic influence
due to the naturalness heuristics will have occurred before any such
result can be established.

Research direction in particle physics moved away from
the considerations of simplicity. This was hardly imaginable even a
short while ago, when, e.g., Quine wrote that ``simplicity, economy
and naturalness\ldots contribute to the molding of scientific
theories generally''~\cite{QuinePTb}. Quine's conventionalist view
is intimately linked with the thesis of empirical underdetermination
of natural science by observable events. It holds that the
acceptance of a theory is a matter of choice guided by
extra-scientific criteria, of which
simplicity is one~\cite{QuineErkenntis,BenM}.

Contrary to Quine, naturalness and
simplicity are frequent rivals and pull physics in different
directions. Dirac believed that in this case beauty has precedence
over simplicity:
\begin{quote}
The research worker, in his efforts to express the laws of Nature in
mathematical form, should strive mainly for mathematical beauty. He
should still take simplicity into consideration in a subordinate way
to beauty.\ldots It often happens that the requirements of
simplicity and beauty are the same, but when they clash the latter
must take precedence~\cite{dirac1939a}.
\end{quote}
Clashes happen more often these days, and the lack of simplicity can
become dramatic. Complexity of some BSM models makes them less
comprehensible, more difficult for doing calculations, and brings
them closer to the status of a theory that we only believe, but do
not know, to exist. Yet the beauty and elegance of simpler and
easier to grasp models is accompanied by their low inverse
fine-tuning probability. How will the rivalry between simplicity and
naturalness end? Many a human researcher find it repulsive enough to
look for less tuned, but also more complex theories which are harder
to describe. Perhaps the difficulty of working with such models and
extracting from them unambiguous predictions suggests a rapid end of
the naturalness-based heuristic, as physicists will seek for
dramatically different, but in a new way simpler solutions.

To this day, naturalness as heuristics has mainly served to support
the claim for the inadequacy of the original Popper's
falsificationism. If we are now concerned with the role of
metaphysical and aesthetic arguments in science, in the future the
greater influence of simplicity may yet prevail on the heuristics of
naturalness. The latter would then be reduced to the purely
circumstantial desire of certain scientists for a self-justification
of their continuing work on the semi-dead physical models.

\singlespacing\footnotesize

\end{document}